\documentclass[12pt]{article}

\newcommand{\x}{arXiv:}
\newcommand{\m}{\mathrm}
\newcommand{\be}{\begin{equation}}
\newcommand{\ee}{\end{equation}}
\newcommand{\ba}{\begin{eqnarray}}
\newcommand{\ea}{\end{eqnarray}}

\usepackage{graphicx}
\usepackage{amssymb}
\usepackage{amsmath}
\usepackage[T1]{fontenc} 
\usepackage[ansinew]{inputenc} 
\usepackage[nosort]{cite}
\newcommand{\inbar}{\vrule height1.57ex width.4pt depth0pt}
\newcommand{\SW}{\relax{\hbox{$\ \inbar\kern-.285em{\rm S}$}}}

\begin{document}
\thispagestyle{empty}
\begin{center}

\null \vskip-1truecm \vskip2truecm

{\Large{\bf \textsf{The Effect of Vorticity on QGP Thermodynamics}}}

{\Large{\bf \textsf{}}}

{\large{\bf \textsf{}}}

\vskip1truecm

{\large \textsf{Brett McInnes
}}

\vskip0.1truecm

\textsf{\\ National
  University of Singapore}
  \vskip1.2truecm
\textsf{email: matmcinn@nus.edu.sg}\\

\end{center}
\vskip1truecm \centerline{\textsf{ABSTRACT}} \baselineskip=15pt
\medskip

Currently there is intense interest in the properties of the Quark-Gluon Plasma produced in peripheral collisions of heavy ions at various facilities, such as the RHIC. In particular, it is essential to understand the difference between such plasmas and their more readily understood counterparts produced in nearly central collisions. The differences arise primarily from the intense magnetic and vorticity fields generated in the QGP in the peripheral case. It has been argued that the magnetic fields might have a profound effect on QGP thermodynamics. Here we will argue, using a gauge-gravity model incorporating the recently proposed \emph{holographic vorticity bound}, that vorticity also has important consequences for the plasma thermodynamics, in particular, for the entropy density at a given impact energy. A crucial point in our analysis is the need to determine the fate of bulk gravitational parameters when the duality translates them to the boundary.

\newpage

\addtocounter{section}{1}
\section* {\large{\textsf{1. The Thermodynamics of ``Peripheral Plasmas''}}}
It is well known that, in its earliest stages of development, the Quark-Gluon Plasma produced in peripheral heavy-ion collisions is permeated by huge magnetic fields \cite{kn:skokov,kn:denghuang,kn:review,kn:tuchin,kn:gergely,kn:magnet,kn:hattori}. Whether these fields can persist into later stages is much debated: for example, see \cite{kn:muller} for a recent, very relevant discussion. If they can, then they are expected to have several very important consequences: in particular, they can affect many aspects of the \emph{thermodynamics} of the plasma: see \cite{kn:shub} for a thorough discussion.

The recent confirmation \cite{kn:STARcoll,kn:STARcoll2} by the STAR collaboration \cite{kn:STAR} at RHIC that, as expected on general theoretical grounds \cite{kn:liang,kn:bec,kn:huang}, these same ``peripheral plasmas'' are characterized by very large \emph{vorticities}, prompts a question: does this vorticity likewise modify the plasma thermodynamics? This is a particularly natural question in view of the many analogies and indeed interactions between magnetic and vorticity fields: see for example \cite{kn:87,kn:dash}).

In order to approach this question, we need to know the location, in the quark matter phase diagram, of these ``highly vortical'' plasmas. The experimental answer to this question \cite{kn:STARcoll,kn:STARcoll2} is quite surprising and suggestive.

Perhaps the most remarkable aspect of the new experimental findings on QGP vorticity is that it is observed to be very \emph{small} for plasmas produced in \emph{high} impact energy collisions. It is not observed at all in peripheral collisions studied at the LHC ALICE facility \cite{kn:bed}, and at first it was not seen \cite{kn:abel} in collisions at impact energies $\sqrt{s_{\m{NN}}} = 62.4$ and $200$ GeV studied by the STAR collaboration; only very recently has it been detected, using a far larger data set, at such impact energies \cite{kn:STARcoll2}; the observed vorticities for $\sqrt{s_{\m{NN}}} = 200$ GeV are so small that observing them was a technical tour de force. Conversely, considerably larger vorticities are clearly in evidence \cite{kn:STARcoll} in collisions at much \emph{lower} impact energies, for example around $\sqrt{s_{\m{NN}}} =  19.6,\, 27,\, 39$ GeV. A precise formulation of the observations is that vorticity \emph{decreases} as the ratio $\alpha/\varepsilon$ of the angular momentum and energy densities increases (this ratio increases with impact energy).

The fact that QGP vorticity is most readily investigated at relatively low impact energies has an important implication: the study of vorticity takes us away from the extremely high impact energies of the LHC, and into the domain of the \emph{Beam Energy Scans} \cite{kn:BEStheory} being conducted at RHIC and elsewhere. From a theoretical point of view, it therefore takes us firmly into the domain of large values of the baryonic chemical potential $\mu_B$, and of \emph{strong coupling}, where one has few analytical techniques.

This, then, is the region of the quark matter phase diagram in which vorticity might play a decisive role in the behaviour of the QGP: relatively low temperatures, and relatively high values of $\mu_B$. This is the domain in which the methods of gauge-gravity duality \cite{kn:nat,kn:veron,kn:wolf} (``holography'') may be helpful. Here, the thermal properties of the QGP are modelled using a dual black hole in an asymptotically AdS ```bulk''. In order to construct a holographic model of vorticity, we need the black hole to have a non-zero angular momentum \cite{kn:schalm}; since the baryonic chemical potential (and possibly the magnetic field) are large, this black hole must be given electric and magnetic charges \cite{kn:myers}.

The observed inverse relation between $\alpha/\varepsilon$ and angular velocity can be formulated, in this language, in terms of a question: is it possible for a massive particle orbiting such a black hole to have a large angular momentum (per unit mass) and yet a small angular velocity?

The answer is that this is indeed possible: the relation between angular momentum and angular velocity for particles orbiting a rotating black hole is not simple \cite{kn:gron}). For example, one often studies particles or observers with zero angular momentum near such a black hole, and these certainly need not have zero angular velocity: this is \emph{frame-dragging}. In \cite{kn:93,kn:95} we showed that such considerations impose, through the gauge-gravity duality, an \emph{upper bound} on the possible values of the vorticity $\omega$ for a given plasma energy density $\varepsilon$ and angular momentum density $\alpha$. When the bound is (nearly) saturated, as is often the case according to the observational data, the relationship between $\omega$ and $\alpha/\varepsilon$ is an inverse one, explaining the extreme smallness of the vorticity produced in collisions generating large angular momentum densities, such as those studied in \cite{kn:STARcoll2}. In short, we can expect the vorticity bound to be relevant to understanding the physics of these particular plasmas.

As we have explained, if we wish to pursue the theoretical study of vorticity in the relatively low-energy domain in which it is most readily observed, we need a full account of the holography of the baryonic chemical potential (and perhaps of the magnetic field). We will see that this is not so straightforward as might be thought; there are crucial parameters which need to be fixed if the theory is to be predictive. We will argue that the holographic vorticity bound can be used to yield plausible estimates for these quantities; the holographic model of the QGP can then be used to predict the effect of vorticity on various properties of the QGP.

In particular, we will argue that vorticity has a significant effect on the thermodynamics of the QGP, just as (extremely) strong magnetic fields do in principle \cite{kn:shub}. (In fact, using the latest data, we will argue that vorticity has in practice a \emph{far stronger} effect than the magnetic fields likely to be encountered in the relatively low-energy domain.) This has potentially major consequences for the interpretation of data currently being collected (pertaining, for example, to \emph{jet quenching}, which is thought to be related to aspects of QGP thermodynamics). In order to show this, we will need to compare situations without vorticity with those in which the vorticity is measurable: that is, we need a holographic model in which plasmas generated in central collisions can be compared with their counterparts produced in peripheral collisions.

We begin by constructing the ``holographic dictionary'' with this objective in view.

\addtocounter{section}{1}
\section* {\large{\textsf{2. Holography Including Vorticity, $\mu_B$, and Magnetic Fields}}}
The global polarization of $\Lambda$ and $\overline{\Lambda}$ hyperons \cite{kn:hyper} in heavy ion collisions is an observable that permits the QGP vorticity to be measured; and the STAR collaboration has reported the discovery and measurement of this important property, with ever-increasing precision \cite{kn:abel,kn:STARcoll,kn:STARcoll2}.

We now review the holographic ``dictionary'' relevant to this vorticity, in the context of non-negligible values of the baryonic chemical potential and the magnetic field. Some aspects of this are familiar, but we argue that others have not previously received sufficient attention.

\addtocounter{section}{1}
\subsubsection* {\textsf{2.1. Holographic Basics: Temperature and Entropy}}
The four-dimensional\footnote{It is customary in collision physics to focus on a two-dimensional section through the system, the \emph{reaction plane} or ``$x - z$ plane''. This is particularly appropriate when studying systems with large magnetic fields or angular momenta, the magnetic field vector or angular momentum vector being taken parallel to the $y$ - axis perpendicular to the reaction plane. In effect, the system can be treated as inhabiting a three-dimensional spacetime, in which the physics is dual to a four-dimensional bulk.} dyonic AdS-Kerr-Newman metric \cite{kn:cognola} has the form
\begin{flalign}\label{A}
g(\m{AdSdyKN)} = &- {\Delta_r \over \rho^2}\Bigg[\,\m{d}t \; - \; {a \over \Xi}\m{sin}^2\theta \,\m{d}\phi\Bigg]^2\;+\;{\rho^2 \over \Delta_r}\m{d}r^2\;+\;{\rho^2 \over \Delta_{\theta}}\m{d}\theta^2 \\ \notag \,\,\,\,&+\;{\m{sin}^2\theta \,\Delta_{\theta} \over \rho^2}\Bigg[a\,\m{d}t \; - \;{r^2\,+\,a^2 \over \Xi}\,\m{d}\phi\Bigg]^2,
\end{flalign}
where
\begin{eqnarray}\label{B}
\rho^2& = & r^2\;+\;a^2\m{cos}^2\theta, \nonumber\\
\Delta_r & = & (r^2+a^2)\Big(1 + {r^2\over L^2}\Big) - 2Mr + {Q^2 + P^2\over 4\pi},\nonumber\\
\Delta_{\theta}& = & 1 - {a^2\over L^2} \, \m{cos}^2\theta, \nonumber\\
\Xi & = & 1 - {a^2\over L^2}.
\end{eqnarray}
Here $L$ is the AdS length scale, $a$ is the angular momentum per unit physical mass, and $M, Q$, and $P$ are parameters (with units of length in natural units) related \cite{kn:gibperry} to the physical mass $m$, the electric charge $q$, and the magnetic charge $p$, by
\begin{equation}\label{C}
m\;=\;M/(\ell_{\mathcal{B}}^2\Xi^2), \;\;\;\;\;q\;=\;Q/(\ell_{\mathcal{B}}\Xi),\;\;\;\;\;p\;=P/(\ell_{\mathcal{B}}\Xi),
\end{equation}
where $\ell_{\mathcal{B}}$ is the gravitational length scale in the bulk, to be discussed in detail below. Notice that $L$, whatever meaning it may have on the boundary, is constrained in the bulk by the value of $a$: for the geometry to make sense (that is, if the metric is to have a consistent signature), we must have
\begin{equation}\label{D}
L \;>\; a.
\end{equation}
We will discuss this inequality in more detail later.

The electromagnetic potential form here is given (see \cite{kn:cognola,kn:87}) by
\begin{flalign}\label{E}
A = &\left(-\,{Qr+aP\,\m{cos}\theta\over 4\pi \ell_{\mathcal{B}} \rho^2}\,+{Q\,r_h+aP\over 4\pi \ell_{\mathcal{B}} \left(r_h^2+a^2\right)}\right)\,\m{d}t \\ \notag \,\,\,\,&+\;\left({1\over 4\pi \ell_{\mathcal{B}}\rho^2}\,\left[Qar\,\m{sin}^2\theta+P\,\m{cos}\theta\left\{r^2+a^2\right\}\right]-{P\over 4\pi \ell_{\mathcal{B}}}\right)\,\m{d}\phi,
\end{flalign}
where $r_h$ is the value of the radial coordinate at the event horizon, which is of course related to the other parameters by
\begin{equation}\label{K}
\Delta_r(r_h)\;=\;(r_h^2+a^2)\Big(1 + {r_h^2\over L^2}\Big) - 2Mr_h + {Q^2 + P^2\over 4\pi}\;=\;0.
\end{equation}

The constant terms in equation (\ref{E}) are important: see \cite{kn:87} for the derivation\footnote{As a check that the various constants here are correctly placed, one can integrate (for example) the electric field, given here (from equation (\ref{E})) by
$$E={-1\over 4\pi \ell_{\mathcal{B}}\rho^4}\left[Q\left(r^2-a^2\m{cos}^2\theta\right)+2Pra\,\m{cos}\,\theta\right],$$
against the element of area in this case, ${r^2+a^2 \over \Xi}\,\m{sin}\,\theta\,\m{d}\theta\,\m{d}\phi$ (for any surface defined by a fixed $r$), obtaining finally $Q/(\ell_{\mathcal{B}}\Xi) = q$ (equations (\ref{C})), so the physical charge $q$ (and not the geometric black hole parameter $Q$) does emerge correctly in Gauss' law.}.

The Hawking temperature of the black hole is given \cite{kn:cognola} by
\begin{equation}\label{F}
T\;=\;{r_h \Big(1\,+\,a^2/L^2\,+\,3r_h^2/L^2\,-\,{a^2\,+\,\{Q^2+P^2\}/4\pi \over r_h^2}\Big)\over 4\pi (a^2\,+\,r_h^2)}.
\end{equation}

The entropy of the black hole is proportional to the area of its event horizon, which is equal to $4\pi (r_h^2+a^2)/\Xi$, so we have
\begin{equation}\label{G}
S\;=\;{\pi\left(r_h^2+a^2\right)\over \ell_{\mathcal{B}}^2\Xi}.
\end{equation}
For holographic purposes it is actually more useful to consider the ratio of the black hole's entropy to its physical mass, given (see the first member of equations (\ref{C})) by
\begin{equation}\label{GG}
{S\over m}\;=\;{\pi\Xi\left(r_h^2+a^2\right)\over M}.
\end{equation}

The ``holographic dictionary'' in this case begins as follows.

As always, the Hawking temperature given by (\ref{F}) will be identified with the temperature of the plasma-like matter in the boundary theory, $T_{\infty}$:
\begin{equation}\label{HAWK}
T_{\infty}\;=\;{r_h \Big(1\,+\,a^2/L^2\,+\,3r_h^2/L^2\,-\,{a^2\,+\,\{Q^2+P^2\}/4\pi \over r_h^2}\Big)\over 4\pi (a^2\,+\,r_h^2)}.
\end{equation}

We can identify the entropy per unit mass of the bulk black hole (equation (\ref{GG})) with the ratio of the entropy density $s$ to the energy density $\varepsilon$ of the boundary plasma:
\begin{equation}\label{HH}
{s\over \varepsilon}\;=\;{\pi\Xi\left(r_h^2+a^2\right)\over M}.
\end{equation}
Note that $\ell_{\mathcal{B}}$ does not appear here, just as it does not appear in the expression for the temperature; though it does appear in the expressions for the entropy and the mass of the black hole separately.

We now turn to some less familiar entries in the holographic dictionary.

\addtocounter{section}{1}
\subsubsection* {\textsf{2.2. Angular Momentum Density and Effective Radius of Gyration}}
The parameter $a$ is interpreted on the boundary as the ratio of the angular momentum density of the plasma to its energy density\footnote{In natural units, the two densities have dimensions respectively of fm$^{- 3}$ and fm$^{- 4}$, so $a$ has units of fm.}. These densities can be estimated (using for example \cite{kn:jiang}) and so, for a given collision of heavy ions, we can assign a definite numerical value to $a$. For example, we will later be considering collisions of gold nuclei, observed by the STAR collaboration, at impact energy 200 GeV and centrality around 20$\%$. Using the estimates given in \cite{kn:jiang}, we find that, in this case, $a \approx 82.5$ fm.

We belabour this point a little because we wish to stress that the AdS length scale parameter $L$, while initially defined in the bulk, must have some interpretation, and therefore \emph{some definite value}, determined by the physics of the boundary theory. This is a simple consequence of the fact that the duality translates \emph{all} aspects of the bulk theory to the (precisely equivalent) boundary theory. More specifically, consider the inequality (\ref{D}) given above, which is essential for the mathematical consistency of the bulk theory. From what we have just said, it is clear that, when discussing the QGP produced in collisions of gold nuclei at 200 GeV impact energy and centrality around 20$\%$, this inequality implies that $L$ must be \emph{some definite number} of femtometres, larger than 82.5. Certainly, we cannot simply assign $L$ some value ``for convenience'': we have to determine its value in a physical way\footnote{We note that the problem of understanding the role of $L$ on the boundary arises also in the study of ``holographic complexity''. In \cite{kn:newermyers} (see particularly Section 6) the authors find that ``the complexity explicitly depends on the AdS curvature scale $L$, which has no interpretation in the boundary theory''. The difficulties involved in attempting to eliminate this additional scale are then discussed; interestingly, none of the natural choices leads to an acceptable outcome, though there may be ways to circumvent this. The same issue is also discussed at the end of Section 2.2.1 of \cite{kn:newmyers} and in Section 5 of that work (``$\ldots$ the AdS scale, which is not a quantity that the boundary CFT should know about$\ldots$''). Our attitude here is different: it is our task to \emph{determine} the value of $L$ by elucidating its role in boundary physics, not to try to eliminate it.}.

Another lesson we learn from the inequality (\ref{D}) is that we should expect $L$ to vary from case to case, since $a$ surely will do so; that is, we cannot expect to describe all of the boundary theories in which we are interested by using a single AdS spacetime. (This is not entirely surprising, because, in the AdS/CFT correspondence, $L$ is related to the fundamental field theory parameters, such as the number of colours, through its \emph{ratio} with other quantities having dimensions of length; so it is these ratios that should be fixed, not $L$ itself, if one is considering (say) a set of collisions all at the same impact energy. See Section 2.3 below.)

Fortunately, it is possible to argue that holography itself supplies an approach to determining $L$ in each case that concerns us. This works as follows (see \cite{kn:93,kn:95} for the technical background).

As mentioned above, the fact that the observed QGP vorticities are unexpectedly small suggests that there may be a \emph{bound on QGP vorticity}, and such a bound can indeed be derived from the gauge-gravity duality. To be precise: the duality furnishes a relation between the vorticity $\omega$ and the ratio $a = \alpha/\varepsilon$ of the angular momentum and energy densities, taking the form
\begin{equation}\label{L}
\omega\;=\;{a\over L^2}\,\sqrt{{1 - {a^2\over L^2}\over 1\;+\;{a^2\over L^2}\,\left(1 - {a^2\over L^2}\right)}}\;;
\end{equation}
this is derived by studying the relationship between the angular momentum per unit mass and the angular velocity of particles in orbit around the bulk black hole. As mentioned in Section 1, one expects a non-trivial relationship between the two when the black hole is rotating, and, as can be seen, this is the case here: $\omega$ and $a$ are related linearly only when $a$ is very small relative to $L$.

If we fix $a$ (regarding it as being determined by the impact energy and centrality of a given collision), then $\omega$ can be regarded as a function of $L$, on the domain $(a,\; \infty)$. An example of this is shown in Figure 1, for a collision with impact energy 27 GeV and centrality around $20 \%$.
\begin{figure}[!h]
\centering
\includegraphics[width=0.9\textwidth]{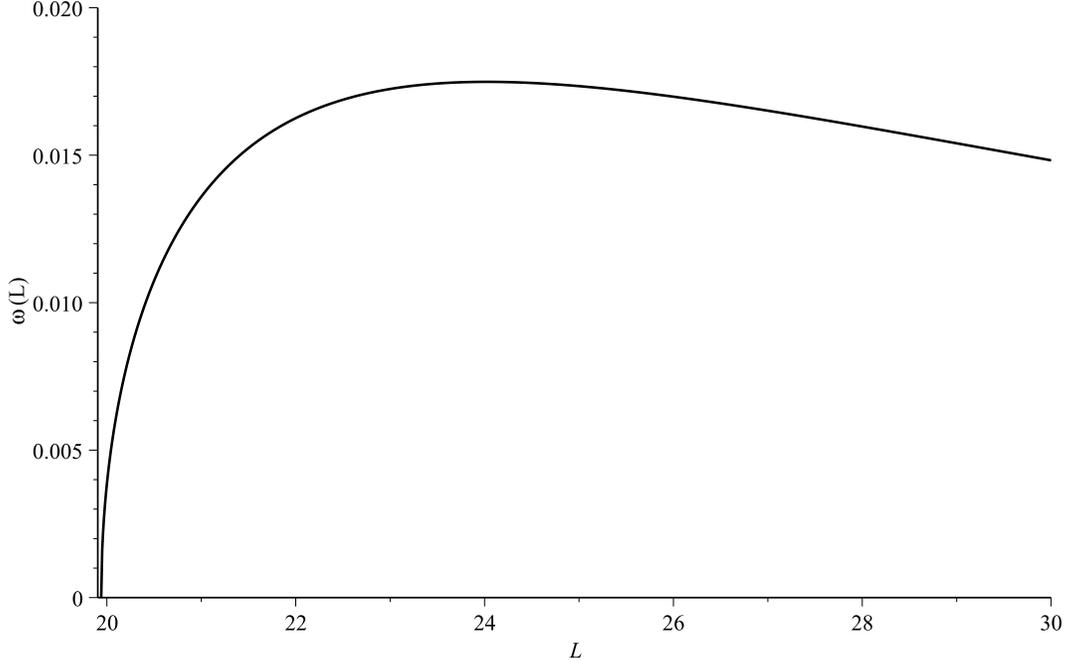}
\caption{Angular velocity (units of fm$^{-1}$) as a function of $L$, with impact energy $\sqrt{s_{\m{NN}}} = $ 27 GeV, centrality $\mathcal{C} = 20 \%$.}
\end{figure}

One sees that indeed $\omega$ is bounded above, whatever the value of $L$ may be. It is straightforward to show that this maximum is inversely proportional to $a$, and, with the holographic interpretation of $a$ as $\alpha/\varepsilon$, this explains why the vorticity is small in collisions with large impact energies (and therefore large values of $\alpha/\varepsilon$).

Now in fact it turns out that, except at very low impact energies, the observed vorticities not only respect this bound: to a good approximation, they \emph{saturate} it. This is the case \cite{kn:93,kn:95} for nearly all\footnote{It is not the case for the lowest impact energy considered in \cite{kn:STARcoll}, namely 7.7 GeV. However, the reported data in that case are rather anomalous, in the sense that the reported difference between the average polarizations of $\Lambda$ and $\overline{\Lambda}$ hyperons is very extreme. This case perhaps requires further attention, both experimental and theoretical (see in this connection the recent interesting observations in \cite{kn:csernkap}).} of the impact energies considered in \cite{kn:STARcoll,kn:STARcoll2}, namely $\sqrt{s_{\m{NN}}} = 11.5,\, 14.5,\, 19.6,\, 27,\, 39,\, 62.4,\, 200$ GeV (at around $20 \%$ centrality). In these cases, then, $L$ must be approximately given by that value, $L_{\omega_{\m{max}}}$, which corresponds to the maximum visible in Figure 1 (and the corresponding graphs for the other impact energies). One can easily show that $L_{\omega_{\m{max}}}$ is a universal dimensionless multiple (denoted $\varsigma$) of $a$:
\begin{equation}\label{LL}
L_{\omega_{\m{max}}}\;=\; \varsigma \,a\, \approx 1.2048\, a.
\end{equation}
This gives us a way of computing $L$ for given collision parameters, in those cases where the vorticity bound is attained. In cases where the bound is not attained, we will obtain $L$ by means of a simple extrapolation: this is needed, for example, to determine $L_0$, the value of $L$ for central collisions: that is, $L_0 \equiv L(a = 0)$.

From a theoretical point of view, one can understand the physical role of $L$ in this application by considering the classical \emph{radius of gyration}, traditionally denoted $k$, defined as the square root of the ratio of the angular momentum per unit mass of any steadily rotating rigid object to its angular velocity. In the present case, we can define an analogous quantity, which we can call the \emph{effective radius of gyration} of a plasma vortex, as $k_{\m{eff}} = \sqrt{a/\omega}$. This is no more than a definition: a plasma sample is not a rigid object, in particular it expands extremely rapidly after the collision, so we do not expect to be able to compute $k_{\m{eff}}$ from simple geometric considerations; one should think of it as an average value over the lifetime of the plasma, meaning that it can be expected to be substantially larger than the value computed from data on gold nuclei (even for central collisions). Nevertheless, this way of thinking about the parameters is useful: we have, from equation (\ref{L}),
\begin{equation}\label{LLL}
k_{\m{eff}}\;=\;L\,\left({1\;+\;{a^2\over L^2}\,\left(1 - {a^2\over L^2}\right)\over 1 - {a^2\over L^2}}\right)^{1/4}\,,
\end{equation}
from which we see at once that $L_0$, the value of $L$ for central collisions, has the following interpretation: \emph{it is the effective radius of gyration for a plasma sample produced in such a collision} ($\alpha = a = 0$).

In the cases in which the vorticity bound is saturated, one can easily compute $k_{\m{eff}}$ using equations (\ref{LL}) and (\ref{LLL}), combined with data determining $a$ (see \cite{kn:93} for the details). The results are shown in Figure 2 for collisions with $\sqrt{s_{\m{NN}}} = 11.5,\, 14.5,\, 19.6,\, 27,\, 39,\, 62.4$ GeV and $20 \%$ centrality. (For clarity, we have omitted the data point (200 GeV, 139.7 fm): it follows the same pattern as seen in the Figure, that is, it lies close to the least-squares regression line.)
\begin{figure}[!h]
\centering
\includegraphics[width=0.8\textwidth]{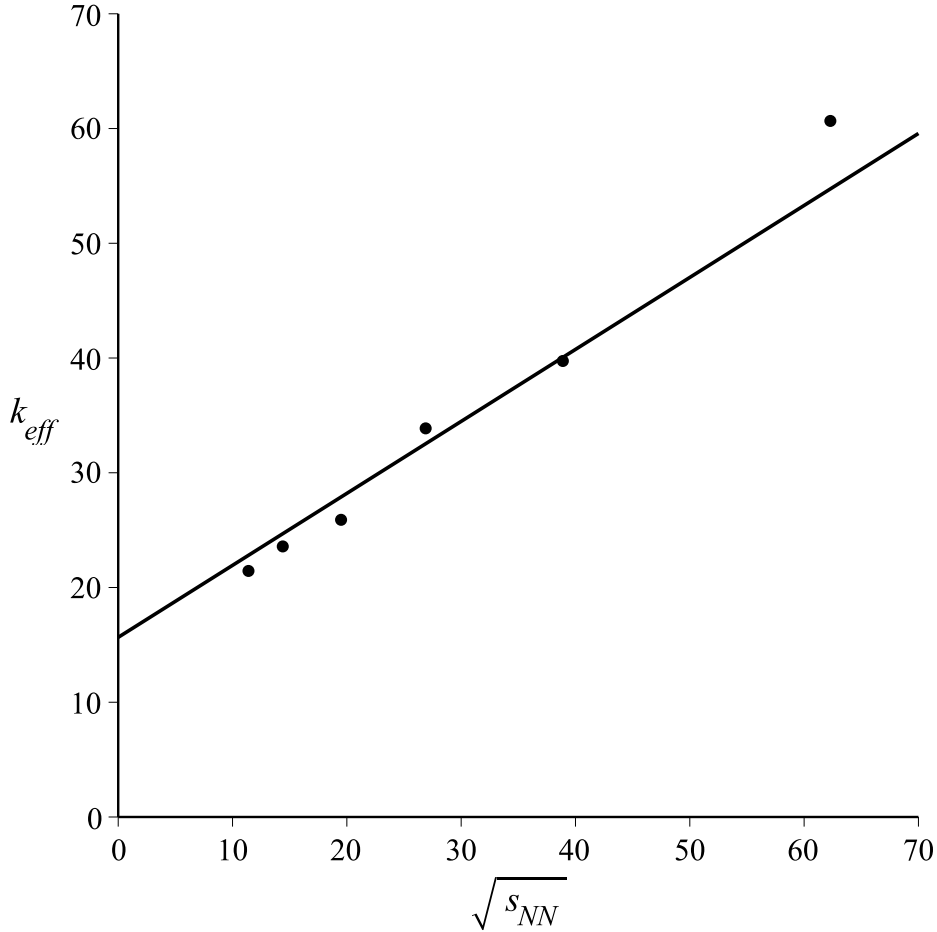}
\caption{$k_{\m{eff}}$ (in femtometres) for gold ion collisions at $\sqrt{s_{\m{NN}}} = 11.5,\, 14.5,\, 19.6,\, 27,\, 39,\, 62.4$ GeV, $20 \%$ centrality. The point (200 GeV, 139.7 fm) has been omitted for clarity; it lies very close to the least-squares regression line (computed including it) shown.}
\end{figure}
The interpretation of Figure 2 is as follows. As the impact energy increases, for collisions at fixed centrality, the angular momentum density likewise increases; in fact, $\alpha$ increases approximately linearly\footnote{The parameter $a$ also increases with impact energy, but not linearly, since the energy density also increases, though considerably more slowly than $\alpha$.} with the impact energy \cite{kn:jiang}. Thus we see that the presence of angular momentum has a strong tendency to \emph{increase} the effective radius of gyration of the plasma vortices, and this may help to explain the smallness of the vorticity in high-energy collisions. A closely related possibility (that impact energy affects the moment of inertia of the plasma forming vortices) was suggested, in general terms, in \cite{kn:jiang,kn:lambdahui2}.

Because of the approximately linear relation between angular momentum density and impact energy, we can think of the horizontal axis as a proxy for the angular momentum density. In view of the fact that the points shown are close to a straight line (as mentioned, the point not shown, (200 GeV, 139.7 fm), also lies very close to this line), we can extrapolate a least-squares regression line to obtain an estimate of $k_{\m{eff}}(\alpha = 0)$, that is, $L_0$: we find $L_0 \approx 15.7$ fm. In view of the discussion above, this is not an unreasonable order-of-magnitude estimate for the rapidly expanding plasmas generated by collisions of heavy ions, for which physical length scales are typically in the 1 - 10 fm range.

To summarize: by using the holographic vorticity bound, we have been able to find interpretations for, and reasonable numerical estimates of, the bulk parameters $a$ and $L$, in terms of boundary physics: they are related to the angular momentum density and to the effective radius of gyration of the system.

\addtocounter{section}{1}
\subsubsection* {\textsf{2.3 Baryonic Chemical Potential and Magnetic Field}}
The final two entries in the ``dictionary'' concern the baryonic chemical potential $\mu_B$ of the plasma, and the magnetic field at infinity, $B_{\infty}$. They require an extended discussion, because they differ from $T_{\infty}$, $s/\varepsilon$, $a$, and $L$, in a simple yet, for us, crucial way: they are obtained by means of a limiting process, taking the radial coordinate $r$ to infinity. In this they are similar to the boundary metric, obtained in the same way; but this means that, like the boundary metric, they are \emph{ambiguous, up to an overall scaling factor}\footnote{Recall that, for example, the magnetic field is extracted from the field strength tensor by expressing the latter in terms of an orthonormal basis. But, to the extent that one has a scaling ambiguity in the boundary metric, so also \emph{orthonormal} bases are ambiguous.}.

Essentially this same point was made, in a somewhat more physical way, in \cite{kn:myers}: see the discussion of ``the relative normalization of the gauge and gravity kinetic terms'' in Section 2.1 of that work (and also its Section 2.3). The point made there is that, with the gravitational/gauge Lagrangian usually employed in the bulk, the electromagnetic potential form has the ``wrong'' units: the timelike component must have units of energy or reciprocal length, but this can only be achieved by introducing an additional length scale, denoted in \cite{kn:myers} as $L_*$.

To relate this to the scaling ambiguity mentioned above, note one has (from \cite{kn:myers}) $L_* \propto \ell_{\mathcal{B}}^{3/2}$ in the five-dimensional case; in the four-dimensional case with which we are concerned in this work, $L_*$ and $\ell_{\mathcal{B}}$ are related by simple proportionality. The dimensionless quantity $L_*/L$ is therefore (in four dimensions) proportional to $\ell_{\mathcal{B}}/L$, which, in holography \cite{kn:nat}, has a definite meaning and a fixed value: one has $\ell_{\mathcal{B}}^2/L^2 = 3/(2N_c)^{3/2}$, where $N_c$ is the number of colours in the boundary field theory\footnote{In this case the bulk dual involves $N_c$ M2-branes (see \cite{kn:AdS4}). This replaces the more familiar relation $\ell_{\mathcal{B}}^3/L^3 = \pi/(2N_c^2)$ for a five-dimensional bulk.}. Thus it makes sense to define a dimensionless quantity $\varpi \equiv L_*/L$, since this quantity is determined for a boundary field theory with a given number of colours and gauge coupling.

We will assume that all plasmas produced in heavy-ion collisions at a given impact energy are described by a single boundary theory with some fixed number of colours and gauge coupling: that is, all such plasmas are described by the same value of $\varpi$, which is fixed in this sense. (Myers et al. also fix $L_*/L$ at a definite value, but they do not attempt to determine what this value might be, selecting $L_*/L = \pi$ (their equation 2.23) simply for convenience.) Here we will take a different point of view: we regard $\varpi$ as a physical parameter \emph{which must have a numerical value fixed by data}\footnote{As we will see, in every case we consider, $\varpi$ does in fact turn out to be of the same order of magnitude as $\pi$; so there is no great disagreement with \cite{kn:myers}.}.

With this understood, we have, taking the limit in equation (\ref{E}) (and in the corresponding field strength 2-form),
\begin{equation}\label{I}
B_{\infty}\,=\,{\Xi \,P\over \varpi L^3}
\end{equation}
and
\begin{equation}\label{J}
\mu_B \,=\,{3\left(Q\,r_h+aP\right)\over 4\pi \varpi L\left(r_h^2+a^2\right)}.
\end{equation}

Notice that $\varpi$ appears only in these two equations. If one is dealing with a plasma with an extremely high temperature, so that $\mu_B$ can be neglected, and if one is interested in plasmas produced in very central collisions, or in phases of the evolution of the plasma in which the magnetic field has attenuated to such an extent that it can be neglected, or if one is simply concerned with issues unrelated to these parameters (as for example in \cite{kn:93,kn:95}), then the precise value of $\varpi$ is not needed. In the situations to be considered in this work, however, the case is very different, since $\mu_B$ certainly, and $B_{\infty}$ possibly, are not negligible if one wishes to investigate the effect of vorticity on thermodynamic variables.

Physical data pertaining to the boundary plasma produced in \emph{central} collisions allow us to specify $T_{\infty}$, $s/\varepsilon$, and $\mu_B$ (as well as, of course, $a = 0, B_{\infty} = 0$): see for example the discussions in \cite{kn:olli,kn:sahoo}. \emph{Since we know $L_0$}, we have now five equations ((\ref{K}), (\ref{HAWK}), (\ref{HH}), (\ref{I}), and (\ref{J})), which can be solved for the black hole geometric parameters $Q, P, M, r_h$, \emph{and also} $\varpi$.

For example, let us consider central collisions at $\sqrt{s_{\m{NN}}} = $ 200 GeV. For consistency, we take the values of $T_{\infty}$, $s/\varepsilon$, and $\mu_B$ in this case from the same source, \cite{kn:sahoo} (where the values of these parameters are indeed given only for central collisions): we have $T_{\infty} \approx$ 190 MeV, $s/\varepsilon \approx$ 1.284 fm (see also \cite{kn:olli}), and $\mu_B \approx$ 30 MeV. Using $L_0 \approx 15.7$ fm and $a = B_{\infty} = 0,$ we solve (\ref{K}), (\ref{HAWK}), (\ref{HH}), (\ref{I}), and (\ref{J}) in this case, and find that $\varpi(\sqrt{s_{\m{NN}}} =  200$ GeV) $\approx 9.177.$ (Collisions at lower impact energies lead to smaller values of $\varpi$ (for reasons explained below), but, in all cases we consider, $\varpi$ lies between 1 and 10.)

In short, we now have estimates of $L$ and $\varpi$ for given physical conditions, and so we are finally in a position to make some predictions. Our main example is as follows.

\addtocounter{section}{1}
\section* {\large{\textsf{3. Case Study: The Effect of Vorticity on Entropy Density}}}
We have seen that, in central collisions with impact energy 200 GeV, the ratio $s/\varepsilon(a = 0, B_{\infty} = 0)$ of the entropy density to the energy density of the QGP is around 1.284 fm. One expects, on general grounds, that this quantity (unlike the temperature or $\mu_B$) could be substantially different for plasmas produced in peripheral collisions: for the strong magnetic fields permeating these ``peripheral plasmas'' might constrict the relevant phase space \cite{kn:shub}, with a consequent strong effect on the entropy density. The question we wish to ask is whether vorticity might have a similar effect, and which of the two is more significant in this sense.

In the 200 GeV impact energy case, we consider centralities $\mathcal{C}$ in the $20 - 50 \%$ range studied in \cite{kn:STARcoll}. For reasons explained in \cite{kn:93}, the parameter $a$ varies very little in this domain, so we use its value at $\mathcal{C} = 20 \%$, which (from \cite{kn:93}) is approximately 82.5 fm. Since the vorticity bound is saturated here \cite{kn:STARcoll2,kn:95}, $L$ can be computed from equation (\ref{LL}). Since we now know $\varpi$, all that remains is to discuss $B_{\infty}$.

As we have stated, there is no doubt that the magnetic field is enormous at the very beginning of the plasma lifetime: for collisions at impact energy 200 GeV and centrality ranging between $20 \%$ and $50 \%$, this initial value probably ranges between 5.2 and 7.7 fm$^{-2}$ (with a strong dependence on centrality) \cite{kn:denghuang}. The question is whether, as a naive estimate based on the departure of the spectator nucleons would suggest, this magnetic field attenuates very rapidly, and so might not have much effect over the lifetime of the plasma.

The most recent observational evidence related to this question comes, very interestingly, from the \emph{same} observations \cite{kn:STARcoll} that uncovered the existence of vorticity resulting from the collisions we have been discussing: M\"{u}ller and Sch\"{a}fer observe \cite{kn:muller} that, while the \emph{sum} of the global polarization fractions of $\Lambda$ and $\overline{\Lambda}$ hyperons is sensitive to the vorticity \cite{kn:hyper}, their \emph{difference} puts a bound on the magnetic field late in the lifetime of the plasma. This bound proves to be very small, suggesting clearly that, in this case at any rate, the magnetic field does attenuate rapidly.

Rather than take a stand on this, we will consider three possible scenarios: we take $B_{\infty}$ to have its maximal possible value in these circumstances, $B_{\infty} = B_{\m{max}} = 7.68$ fm$^{-2}$ (which, in view of \cite{kn:muller}, must now be considered rather unlikely); then, more plausibly, $B_{\infty} = B_{\m{max}}/10$; and, finally, $B_{\infty} = 0$.

Having fixed all of these parameters, we have now only four unknowns ($Q, P, M, r_h$), and they can be found by solving the four equations (\ref{K}), (\ref{HAWK}), (\ref{I}), and (\ref{J}). We can then use equation (\ref{HH}) to \emph{compute} the value of $s/\varepsilon$ for collisions at this impact energy and centrality.

Our results in this case ($\sqrt{s_{\m{NN}}} =  200$ GeV) are as follows.

[1] Leaving aside, for the moment, the effect of the magnetic field, the effect of vorticity is to \emph{reduce} the entropy density to energy density ratio, by a very substantial factor: the value drops from $s/\varepsilon(a = 0, B_{\infty} = 0) \approx 1.284$ fm to $s/\varepsilon(a = 82.5, B_{\infty} = 0) \approx 0.399$ fm. Note that, in the case of magnetic fields \cite{kn:shub}, $\varepsilon$ is \emph{reduced} (only to a small extent, however, for the field values encountered here); if the analogy between vorticity and magnetic fields continues to hold here, then the reduction of $s/\varepsilon$ means that vorticity induces a still larger decrease in the entropy density $s$ itself. Even allowing for the fact that holographic numerical predictions are not always very precise, it is clear that we are entitled to assert that the holographic model predicts that \emph{the vorticity induced in the QGP in peripheral collisions has a very strong effect on the entropy density and therefore on plasma thermodynamics more generally}.

[2] If we now include a magnetic field of the maximal possible strength here, we find that $s/\varepsilon$ declines still more drastically, to $s/\varepsilon(a = 82.5, B_{\infty} = B_{\m{max}}) \approx 0.111$ fm, confirming the claim \cite{kn:shub} that such extremely intense magnetic fields arising in peripheral collisions, were they realistic, would also have a very strong effect on the entropy density. This may also indicate that the mechanism of the reduction is similar in the two cases.

[3] However, the effect of magnetic fields of more realistic magnitudes is much less marked: we find that $s/\varepsilon(a = 82.5, B_{\infty} = B_{\m{max}}/10) \approx 0.356$ fm, only a little lower than $s/\varepsilon(a = 82.5, B_{\infty} = 0)$. In reality, then, it seems that vorticity is by far the dominant factor in reducing the entropy density in this case.

We have repeated these computations for the other impact energies and centralities considered in \cite{kn:STARcoll} where the vorticity bound is approximately attained, that is, for $\sqrt{s_{\m{NN}}} = 11.5,\, 14.5,\, 19.6,\, 27,\, 39,\, 62.4$ GeV, at $20 \%$ centrality. The results are shown in the table; recall that $s/\varepsilon$ has units of femtometres (1 fm $\approx (197.327$ MeV)$^{-1}$).
\begin{center}
\begin{tabular}{|c|c|c|c|c|}
  \hline
$\sqrt{s_{\m{NN}}}$ (GeV) & $s/\varepsilon(0, 0)$   & $s/\varepsilon(a, 0)$ & $s/\varepsilon(a, B_{\m{max}}/10)$ & $s/\varepsilon(a, B_{\m{max}})$ \\
\hline
$11.5$ &  1.344  & 0.418 & 0.418 & 0.418\\
$14.5$ &  1.341  & 0.417 & 0.417 & 0.417\\
$19.6$ &  1.335  & 0.415 & 0.415 & 0.414\\
$27$  &  1.335 & 0.415 & 0.415 & 0.410\\
$39$  &  1.320  & 0.411  & 0.411 & 0.392\\
$62.4$  &  1.317  & 0.410  & 0.408 & 0.319\\
$200$  &  1.284  & 0.399  & 0.356 & 0.111\\
\hline
\end{tabular}
\end{center}
The results are similar in these cases ---$\,$ indeed, to three decimal places, the ratio $[s/\varepsilon(a, 0)]$/$[s/\varepsilon(0, 0)]$ has the \emph{same} value, around 0.311, in every case ---$\,$ except that, at the lower impact energies, the magnetic field has even less effect than at $\sqrt{s_{\m{NN}}} =  200$ GeV. Holographically, this is to be understood as follows: at low impact energies, the magnetic field is smaller in absolute terms, while $\mu_B$ is very much larger (it increases from around 30 MeV in $\sqrt{s_{\m{NN}}} =  200$ GeV collisions up to around 290 MeV when $\sqrt{s_{\m{NN}}} =  11.5$ GeV). This means that the dimensionless quantity $B_{\infty}/\mu_B^2$ is very much smaller for low-energy collisions than for their high-energy counterparts. However, in the holographic model, $B_{\infty}/\mu_B^2$ is proportional to the parameter $\varpi$, so we expect this parameter to be smaller for low-energy collisions, and this is certainly the case in all of the examples we have studied. However, when this $\varpi$ is small, $B_{\infty}$ is, effectively, reduced still further (as can be seen by transferring $\varpi$ to the left side of equation (\ref{I})), and so it has little influence on the system.

In short, in the domain in which vorticity is most prominent (particularly when $\sqrt{s_{\m{NN}}} = 19.6,\, 27,\, 39$ GeV), the magnetic field has almost no impact on the entropy density. (Conversely, the magnetic field dominates over vorticity in very high energy collisions, where, in fact, vorticity is undetectably small; that is, the findings of \cite{kn:shub} are relevant for heavy-ion collisions at the LHC, but not for the RHIC collisions.)

In summary, then, for those collisions in which vorticity is detectable, the holographic model predicts that the vorticity greatly reduces the ratio $s/\varepsilon$, to around one third of its value in the corresponding central collisions; this probably means that $s$ is reduced by about this factor, or perhaps slightly more. The effect is indeed due to the vorticity, not to the magnetic field.

\addtocounter{section}{1}
\section* {\large{\textsf{4. Conclusion}}}
The recent direct observations of QGP vorticity \cite{kn:STARcoll,kn:STARcoll2} have opened a new line of investigation. The question is whether ``highly vortical'' plasmas differ substantially from their better-known counterparts generated by nearly central collisions.

We have argued here, using a holographic or gauge-gravity model, that the vortical plasmas do differ very substantially from the non-vortical QGP: in particular, they are characterised by a far smaller entropy density to energy density ratio. This has many ramifications.

For example, consider the effect of vorticity on the diffusion of momentum in the QGP. This is measured by the \emph{kinematic viscosity} $\nu$, which is the ratio of the dynamic viscosity $\eta$ to the energy density: this is the parameter that occurs in the Navier-Stokes equations. In a holographic model based on Einstein gravity (only), as is the case here, the ratio $\eta/s$ is fixed\footnote{It does of course vary, in reality, with respect to variations of temperature and baryonic chemical potential \cite{kn:bassagain,kn:QGPparameters}, but this variation is not significant in the kind of comparisons we are making here.} \cite{kn:nat}; but then since we have
\begin{equation}\label{U}
\nu\;=\;{\eta\over \varepsilon}\;=\;{\eta\over s}\times{s\over \varepsilon},
\end{equation}
this means that $\nu$ is significantly smaller in the vortical plasma than in the central case. Thus the flow of the QGP ``liquid'' might well be strongly affected by vorticity.

Again, the much-studied jet quenching parameter $\hat{q}$ is related, in holographic models \cite{kn:hong1,kn:hong2}, to the entropy density, and so jet quenching might also have unusual aspects in holographic models of the QGP, particularly for collisions at relatively low energies \cite{kn:beamenergyjet}. We will return to this important question elsewhere.

\addtocounter{section}{1}
\section*{\large{\textsf{Acknowledgements}}}
The author thanks Dr Soon Wanmei for valuable discussions.

\end{document}